\documentclass{aa}  
\usepackage{hyperref}
\usepackage{textcomp}
\usepackage{graphicx} %
\usepackage{amsmath} %
\usepackage{amssymb} %
\usepackage{bm} %
\usepackage{upgreek} %
\usepackage{IEEEtrantools} %
\usepackage{multirow}
\usepackage{txfonts}
\usepackage[normalem]{ulem} 
\usepackage{multirow}
\usepackage[dvipsnames]{xcolor}
\usepackage{IEEEtrantools}

\newcommand{\sect}[1]{\text{Sect.~\ref{#1}}}
\newcommand{\fig}[1]{\text{Fig.~\ref{#1}}}

\newcommand{\kmos}{\mathrm{km\,s^{-1}}}

\newcommand{\nm}{\mathrm{nm}}
\newcommand{\dex}{\mathrm{dex}}
\newcommand{\teff}{T_{\mathrm{eff}}}
\newcommand{\lgg}{\log\,g}
\newcommand{\lggeff}{\log\,g_{\mathrm{eff}}}
\newcommand{\trip}{\ion{O}{I} $777\,\nm$}
\newcommand{\lowerlev}{$3\mathrm{s}^{5}\,\mathrm{S}$}
\newcommand{\upperlev}{$3\mathrm{p}^{5}\,\mathrm{P}$}

\begin{document} 
        
        \title{Two-dimensional non-LTE \ion{O}{I} 777\,nm line formation in radiation hydrodynamics simulations of  Cepheid atmospheres}
        
        \subtitle{}
        
        \author{V. Vasilyev \inst{1,2}, A. M. Amarsi \inst{3}, 
                H.-G. Ludwig \inst{2}, and B. Lemasle \inst{4}}
        
        \institute{ Max-Planck-Institut f\"ur Sonnensystemforschung, 
                Justus-von-Liebig-Weg 3, 37077 G\"ottingen, Germany \\
                \email{vasilyev@mps.mpg.de}              \and  
                Landesternwarte, Zentrum für Astronomie der Universität Heidelberg, Königstuhl 12, 69117 Heidelberg, Germany \\ 
                \email{hludwig@lsw.uni-heidelberg.de}    \and 
                Max Planck Institute f\"ur Astronomy,
                K\"onigstuhl 17, D-69117 Heidelberg, Germany  \\
                \email{amarsi@mpia.de}                    \and 
                Astronomisches Rechen-Institut, Zentrum für Astronomie der Universität Heidelberg, Mönchhofstr. 12-14, D-69120 Heidelberg, Germany \\
                \email{lemasle@uni-heidelberg.de}}
        \date{Received ; accepted }
        \titlerunning{2D non-LTE \ion{O}{I} 777\,nm line formation in RHD simulations  of Cepheid atmospheres}
        \authorrunning{V.~Vasilyev et al.}  
        
        \abstract{ Oxygen abundance measurements are important for
          understanding stellar structure and evolution.  Measured in
          Cepheids, they further provide clues on the metallicity gradient and
          chemo-dynamical evolution in the Galaxy.  However, most of the
          abundance analyses of Cepheids to date have been based on
          one-dimensional (1D) hydrostatic model atmospheres.  Here, we test
          the validity of this approach for the key oxygen abundance
          diagnostic, the \ion{O}{I} $777\,\mathrm{nm}$~triplet lines. We
          carry out 2D non-LTE radiative transfer calculations across two
          different 2D radiation hydrodynamics simulations of Cepheid
          atmospheres, having stellar parameters of $\teff= 5600$ K,
          solar chemical compositions, and $\lgg= 1.5$ and $2.0$,
          corresponding to pulsation periods of 9 and 3 days, respectively. We
          find that the 2D non-LTE versus 1D LTE abundance differences range
          from $-1.0$~dex to $-0.25$~dex depending on pulsational phase. The 2D
          non-LTE versus 1D non-LTE abundance differences range from $-0.2$~dex
          to $0.8$~dex.  The abundance differences are smallest when the Cepheid
          atmospheres are closest to hydrostatic equilibrium, corresponding to
          phases of around $0.3$ to $0.8$, and we recommend these phases for
          observers deriving the oxygen abundance from \ion{O}{I}
          $777\,\mathrm{nm}$ triplet with 1D hydrostatic models.}
        
        
        \titlerunning{2D non-LTE \ion{O}{I} 777\,nm line formation in RHD simulations  of  Cepheids}
        \authorrunning{V. Vasilyev et al.}  
        \maketitle

        \section{Introduction}
        Oxygen is the third most abundant chemical element and the most abundant metal
        in the Universe.  As with other $\upalpha$-elements, it forms primarily in massive
        stars. As such, oxygen abundances in the atmospheres of stars are important
        tracers of the Galactic chemical evolution \citep[GCE;
        e.g.][]{2003MNRAS.339...63C, 2013A&A...558A...9M, 2015A&A...580A.126K,
                2015A&A...580A.127K}. In particular, they provide strong constraints on the
        formation timescales of Galactic stellar populations (for details, see
                e.g. \citealt{2007A&A...465..271R,2013ApJ...764...78R} and
                \citealt{2017A&A...603A...2M}). 
        
        In this context, classical Cepheids are interesting; as they are young
        \citep[<250 Myr; e.g.][]{2005ApJ...621..966B}, massive, evolved
        stars.          \cite{1998PASJ...50..629T} and later  \cite{2013MNRAS.432..769T} investigated
        the nature of evolution-induced mixing in the envelope of evolved
        intermediate-mass stars in supergiants and Cepheid variables of various
        pulsation periods to determine the photospheric abundances of C, N, O, and Na,
        applying non-LTE analyses. They showed that the observed CNO 
        abundance trends are mainly a result of canonical dredge-up 
        of CN-cycled material, while any significant
        non-canonical deep mixing of ON-cycled gas is ruled out.
        
        With the exception of a few elements whose composition is modified by
        the first dredge-up (Li, C, and N in particular), the chemical
        composition of Cepheids reflects the current chemical composition of
        the interstellar medium (ISM).  Furthermore Cepheids are bright stars
        that can be detected out to large distances. They can be used as
        standard candles for distance measurements due to the existence of a
        relation between their period of pulsation and intrinsic luminosity
        \citep{1908AnHar..60...87L, 1912HarCi.173....1L}.  Combining
        information on the oxygen abundance in Cepheids with their distances,
        we can derive the present-day oxygen abundance gradient of the
        Galactic disc, which is a crucial constraint for the models of the
        Galactic chemical evolution. In this respect, Cepheids 
        adequately complement large-scale surveys such as GALAH
        \citep{2015MNRAS.449.2604D}, APOGEE \citep{2017AJ....154...94M}, and
        GAIA-ESO \citep{2012Msngr.147...25G}, focusing mostly on older stars;
        they are therefore useful for deriving the time evolution of Milky Way
        gradients \citep[e.g.][]{2017A&A...600A..70A}.
        
        A number of studies were performed
        \citep[e.g.][]{2008A&A...490..613L, 2013A&A...558A..31L,
                        2011AJ....142...51L, 2011AJ....142..136L, 2015A&A...580A..17G,
                        2016A&A...586A.125D} to determine Galactic abundance gradients for numerous
        {$\upalpha$}-, 
        iron-peak, and neutron-capture elements using a large number of classical
        Cepheids. However, the abundance measurements presented in these
        Cepheid studies are all based on 1D hydrostatic stellar model atmospheres. In
        these models, convection is typically treated in the framework of the mixing length
        theory \citep{1958ZA.....46..108B}.  Abundances are typically measured at
        different phases, and thereby represent the dynamical, evolving atmosphere as a
        sequence of quasi-hydrostatic states with different effective temperatures and surface
        gravities.  Furthermore, in most of the above-mentioned studies, the
        line formation was assumed to take place under the condition of local thermodynamic
                equilibrium (LTE), although later studies relaxed this assumption and were
        based on non-LTE radiative transfer.
        
        Possible non-LTE effects in Cepheids were discussed by
                \cite{2011AJ....142..136L}. Later, \cite{1538-3881-146-1-18} and
                \cite{2014MNRAS.444.3301K} presented 1D non-LTE oxygen abundance analyses of
                large samples of Cepheids.  The \trip~and
                $845\,\nm$~triplets can be used  to derive oxygen abundances because they are easily
        detectable over a wide range of effective temperature and generally do not
        suffer from severe blending with other species; furthermore, the spectral region
        is less contaminated by terrestrial bands compared to the forbidden
        low-excitation  [\ion{O}{I}] $630.0\,\nm$ and $636.4\,\nm$~lines, or
        to the permitted high-excitation \ion{O}{I} $615.8\,\nm$~triplet.
        However, the near-IR triplets are sensitive to departures from LTE. For a given
        oxygen abundance, taking into account non-LTE effects tends to make these lines
        stronger; thus, the inferred oxygen abundance is lower once the
        assumption of LTE is relaxed. 
        
%
        Since calculations of multi-dimensional Cepheid models are a sizable
        computational problem due to the presence of the different spatial and
        temporal scales, it is not a surprise that the application of 1D
        hydrostatic stellar model atmospheres is common. However, it is
        desirable to obtain an understanding of the limitations inherent to
        the standard approach.  Working towards this goal, in a previous paper
        \citep{2017A&A...606A.140V} we presented a two-dimensional (2D) model 
        of a short-periodic Cepheid-like variable. The model  has realistic convection  based on  the conservation laws.  Additionally, the  model shows self-excited fundamental mode pulsations  due to the $\kappa$-mechanism \citep{1917Obs....40..290E, 1963ARA&A...1..367Z}.        In the present paper,
        performing similar 2D radiation-hydrodynamics (RHD) simulations we aim
        to investigate non-LTE effects on oxygen, their phase-dependence, and biases in the derived oxygen abundances  compared to a
        standard 1D non-LTE analysis based on hydrostatic
        models.  
         
        Specifically, in Section~\ref{methods} we describe the line
        formation code, atomic model, as well as 1D and 2D model
        atmospheres. This study focuses on one of the key oxygen abundance
        diagnostics, namely the \trip~triplet. The results including line
        profiles, abundance errors and biases of the standard approach are
        presented in Section~\ref{results}. In Section~\ref{discussion} we
        discuss the optimal phase for observers in order to get unbiased
        measurements of the oxygen abundance in Cepheids. We also compare
        the behaviour of departures from LTE in our 2D and 1D model atmospheres.
        
        \section{Methods} 
        \label{methods}
        
        \subsection{2D RHD simulations of Cepheid variables}
        \label{methods_atmos}
        
        \begin{figure*}
                \includegraphics[width=9cm, scale=0.7] {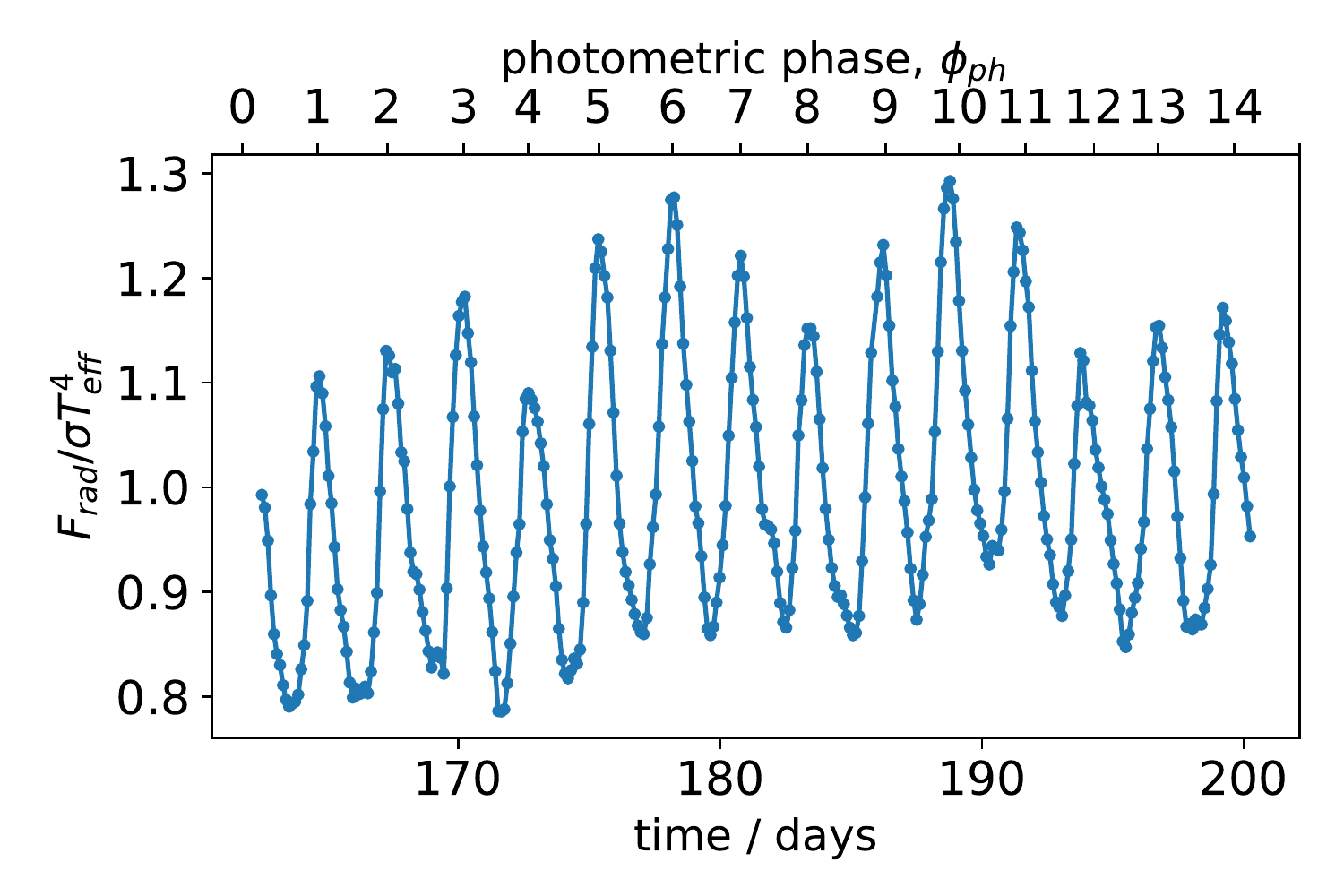}
            \includegraphics[width=9cm, scale=0.7] {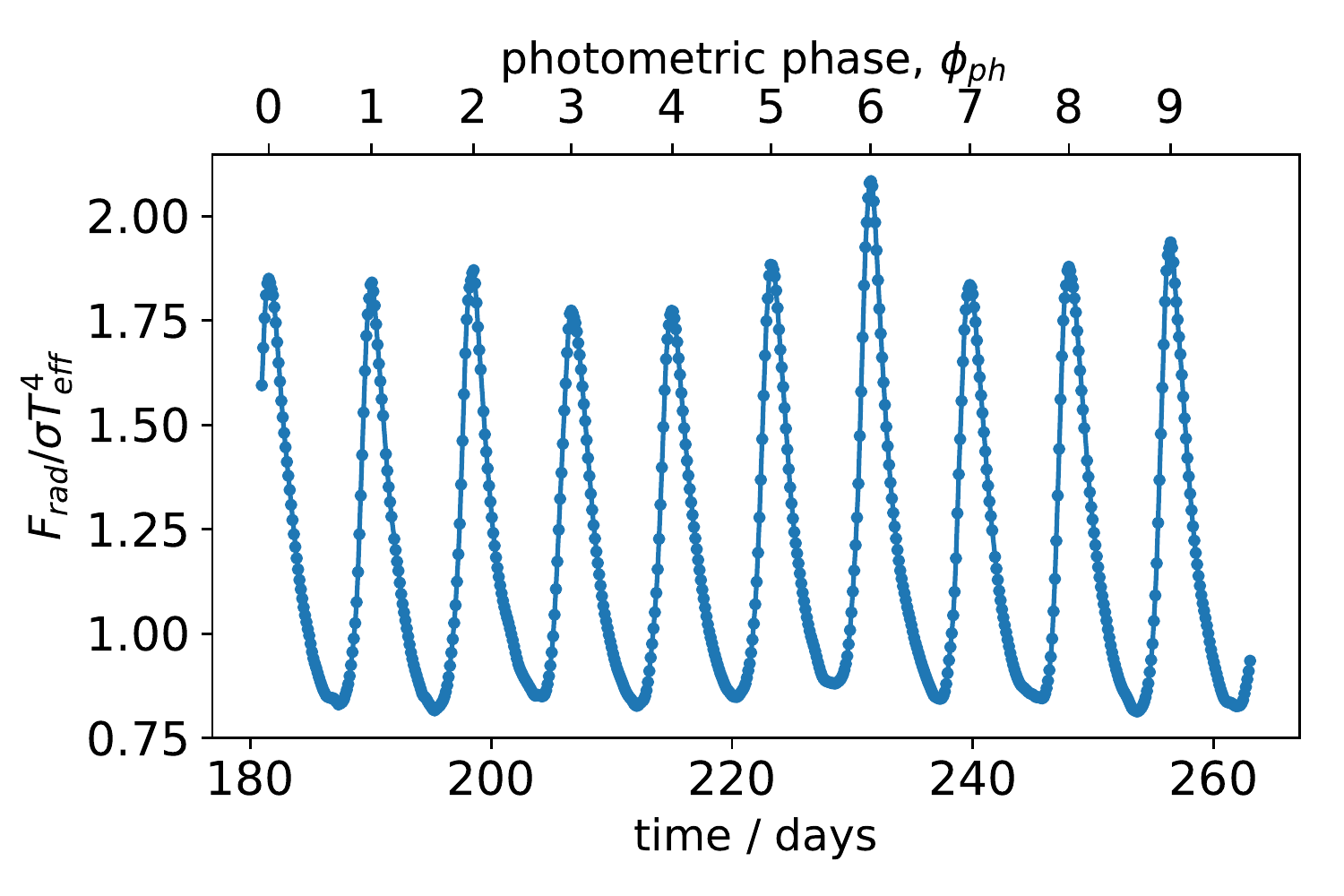}
            \caption{Light curves in terms of the emergent bolometric flux of
              models M3 (left panel) and M9 (right panel) which exhibit
              pulsational periods of 3 and 9 days, respectively.  Blue dots
              indicate the instances in time for which 2D non-LTE spectral
              syntheses were performed. }
                \label{LCgt56g}%
        \end{figure*}
        
        \begin{figure*}
                \includegraphics[width=9cm, scale=0.7] {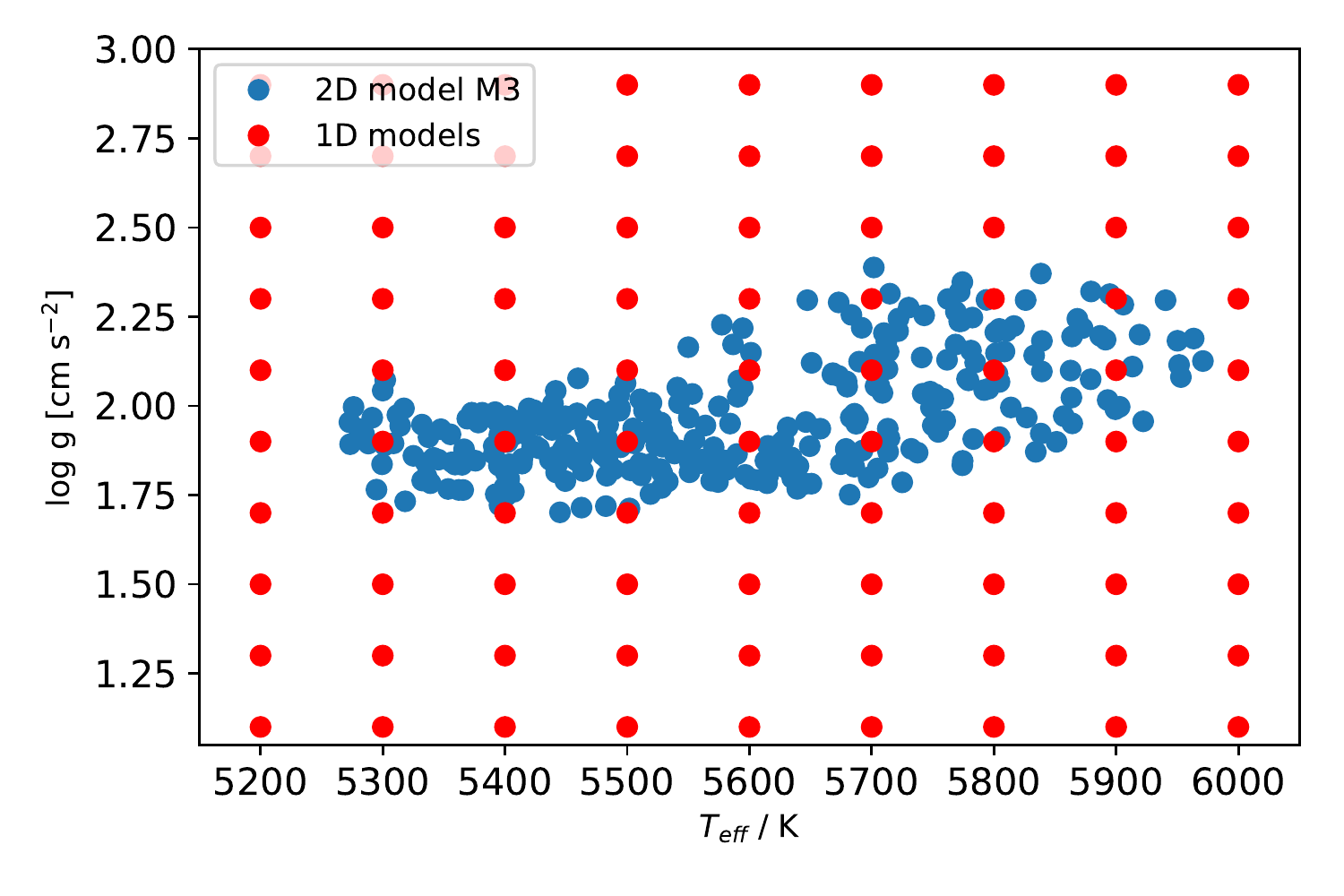}
        \includegraphics[width=9cm, scale=0.7] {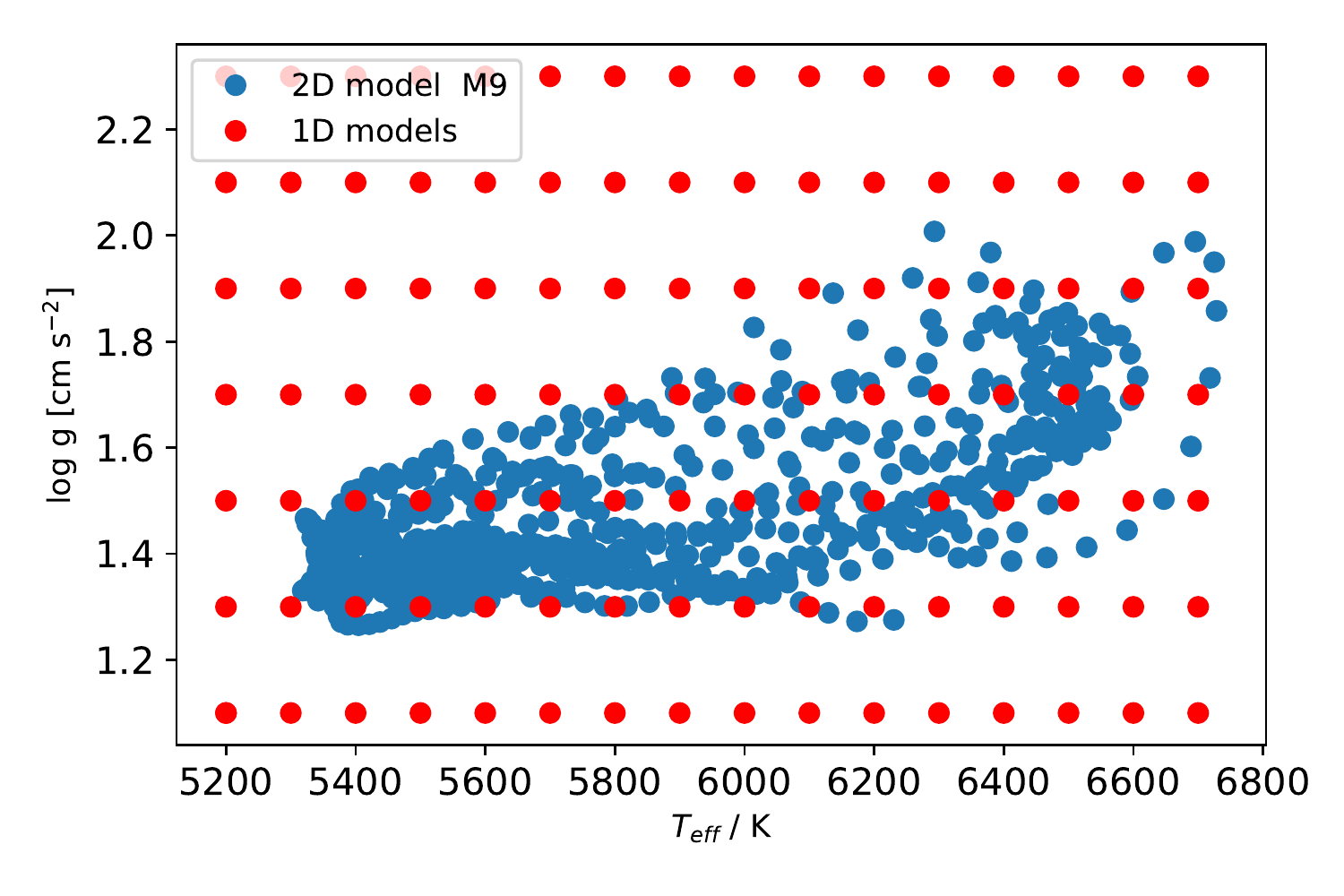}
                \caption{Grid of 1D hydrostatic models (red dots) 
                        in the $\lgg$---$\teff$~plane.
                        The 2D snapshots defined individually by 
                        $\lggeff$ and $\teff$
                        for model M3 (left panel)  and M9 (right panel) are indicated by blue dots.}
                \label{1d_grid}%
        \end{figure*}   

        We performed 2D RHD simulations of Cepheid-like variables using the
        RHD code \textsc{CO5BOLD} \citep{2012JCoPh.231..919F}. The code solves
        the set of 2D compressible RHD equations using a
        short-characteristics method for the radiative transfer, which was
        considered in grey approximation.  Grey \textsc{PHOENIX}
        \citep{2012EAS....57....3A} merged with \textsc{OPAL}
        \citep{1992ApJ...397..717I} opacities were used in the simulations.
        In this study two models are considered:  a three-day model (M3) and a nine-day
        model (M9).  The models have a mean effective
        temperature $\teff=5600\,\mathrm{K}$, constant depth-independent
        gravitational acceleration, $\lgg=1.5$ for M9 and $\lgg=2.0$ for M3,
        and solar-like chemical composition. In \citet{2017A&A...606A.140V} we
        discussed the motivation and limitations of our approach.
        
        The model {M3} has a three-day period of pulsation; it was introduced and
        analysed in \citet{2017A&A...606A.140V}.  In the present study we extended
        the sequence to $14$~full pulsation periods, compared to 6 in our
        previous paper.  We illustrate the bolometric light curve of the model
        in the left panel of Fig.~\ref{LCgt56g}.  Due to the impact of the
        stochastic convective component, the light curve shows cycle-to-cycle
        variations \citep[for details, see][]{2017A&A...606A.140V}.
        
        After relaxation of initial perturbations, model {M9} reaches an
        approximate limit cycle exhibiting fundamental-mode oscillations
        with a period of approximately $9$~days.  We illustrate the bolometric
        light curve of the model in the right panel of Fig.~\ref{LCgt56g}. Due
        to the impact of convective noise, the effective temperature of the
        phases of maximum light varies between $6460\,\mathrm{K}$~and
        $6730\,\mathrm{K}$; in particular, during the seventh cycle the maximum
         effective temperature reaches a value above $6700\,\mathrm{K}$.
        
        While the 2D models have a depth-independent gravitational
        acceleration $\vec{g_\mathrm{o}}$, spectral lines are in fact
        sensitive to the effective gravitational acceleration,
        $\vec{g}_\mathrm{eff}=\vec{g_\mathrm{o}}-\frac{d\vec{v}}{dt}$, which is
        balanced by the pressure gradient.  For the maximum of the seventh
        cycle of the M9 model, the effective gravitational acceleration in the
        photosphere drops by a factor of around three due to the impact of the
        convective noise.  This produces a loop-like structure in our results
        for the abundance corrections and abundance errors, as we 
        discuss in detail in \sect{results}.

        Both simulations use an equidistant spatial Cartesian grid of
        $N_x \times N_z = 600 \times 500$ along the horizontal and vertical
        directions. The geometrical extensions of the models along the
        horizontal and vertical directions are
        $l_\mathrm{x} \times l_\mathrm{z}=1.1\cdot10^{12} \times
        0.5\cdot10^{12}$ cm and
        $l_\mathrm{x} \times l_\mathrm{z}=3.6\cdot10^{12} \times
        1.7\cdot10^{12}$ cm for {M3} and {M9}, respectively.  The models
        both have periodic horizontal boundary conditions, and closed top and
        bottom boundaries.  The pulsations induce strong perturbations in the
        vicinity of the upper (closed) boundary.  In order to damp these
        perturbations a drag force was applied on the horizontal mean motion
        in the uppermost few grid layers.  The impact of this approach on
        spectroscopic properties is discussed in
        \citet[][]{2017A&A...606A.140V}.

        The effective temperature of the 2D models $\teff^\mathrm{2D}$ {was
          calculated from the} bolometric radiative flux passing the top
        boundary at each instance in time.  The kinematic acceleration
        $\frac{d\vec{v}}{dt}$, which is variable during the pulsational cycle,
        leads to a variation in $\vec{g}_\mathrm{eff}$ in the range
        $\approx(0.6 \ldots3.5)\vec{g_\mathrm{o}}$~{for M9, and
          $\approx(0.5 \ldots3.0)\vec{g_\mathrm{o}}$~ for M3}. It was
        calculated from the motion of a (Lagrangian) mass element located
        close to optical depth unity.  We illustrate the variation in $\teff$
        and $\lggeff$ over the pulsational cycles for models M3 and M9 
        in \fig{1d_grid}.

        \subsection{One-dimensional models}
        \label{methods_1d}
                
        {We constructed a grid of 1D hydrostatic model stellar atmospheres
          using the  Lagrangian hydrodynamics (LHD) code}, which
        solves the set of 1D RHD equations in the Lagrangian frame.  In the
        present project we operated it in hydrostatic mode, which  provides
        structures in hydrostatic equilibrium.  {To facilitate a differential
          comparison with the 2D models, the same} opacities and
        equation-of-state tables were used as in the 2D models.  The
        {dimensionless}~mixing-length parameter $\alpha$ was fixed to
        $1.5$~for all 1D models.  In order to cover the parameters encountered
        in M3 and M9, $\teff$, and $ \lgg$ for the 1D models were chosen in the
        range $5200<\teff/\mathrm{K}<6800$ with a step
        $\Delta \teff=100\,\mathrm{K}$ and $1.1<\lgg<2.9$ with
        $\Delta \lgg=0.2$, respectively.  {We depict the grid of 1D models in
          the $\lgg$---$\teff$~plane in \fig{1d_grid} which also shows the
          parameters for the model M3 and M9.}  Technical details concerning
        1D hydrostatic models can also be found in \cite{2018A&A...611A..19V}.

        \subsection{Microturbulent velocity}
        \label{methods_vmic}
        {To calculate 2D LTE/non-LTE versus 1D LTE/non-LTE abundance
          corrections, it is necessary to  first consider the microturbulent velocity
          that needs to be adopted in the spectral synthesis based on 1D model
          atmospheres.}  The microturbulent velocity is a relevant parameter
        in spectroscopic analyses of late-type stars (both pulsating and
        non-pulsating) that are based on 1D hydrostatic models.  It is needed
        to account for the additional line broadening induced by gradients in
        the (convective) velocity fields over distances
        $\Delta \tau_\mathrm{R} \lesssim 1$ {\citep[for details, see
          e.g.][]{2013MSAIS..24...37S}}.  Due to the change in the thermal
        structure and velocity gradients during the pulsational cycle, the
        microturbulent velocity is phase-dependent (for details, see
        {\citealt{2004AJ....128..343L}, \citealt{2005AJ....130.1880A}, and
          \citealt{2005AJ....129..433K}, and the later studies of
          \citealt{2017A&A...606A.140V} and \citealt{2018arXiv180500727P}}).

        {For each 2D snapshot, we calculated an effective microturbulent
          velocity, which we used in the 1D line formation calculations
          (\sect{methods_nlte}) for a comparison of 2D and 1D results.}
        Following \citet{2013MSAIS..24...37S} and \citet{2017A&A...606A.140V},
        these microturbulent velocities were inferred from 2D (LTE/non-LTE)
        spectra.  The microturbulent velocity in the LTE case is calculated by
        equating equivalent widths
        $W(\vec{v}_\mathrm{2D})= W(\xi_\mathrm{t})$, where
        $W(\vec{v}_\mathrm{2D})$ is calculated with the full 2D velocity field
        $\vec{v}_\mathrm{2D}$, and $W(\xi_\mathrm{t})$ by replacing
        $\vec{v}_\mathrm{2D}$ by a depth-independent isotropic microturbulent
        velocity $\xi_\mathrm{t}$.

        As in the case of observations, \ion{Fe}{i} and \ion{Fe}{ii} lines
        were used for the diagnostic of the microturbulent velocity. 
        {The line list, which was given in \citet{2018A&A...611A..19V}, 
        {contains} artificial neutral and singly ionised iron lines with
        excitation potentials $1$, $3$, and $5\,\mathrm{eV}$.}
         Oscillator strengths were chosen
        to cover the linear and non-linear parts of the curve of growth.  The
        spectral synthesis was performed using the code {\textsc{LINFOR3D}
          \citep{2017MmSAI..88...82G}}.  Measured microturbulent velocities
        were averaged over line strengths and excitation potentials within
        ionisation specie for each individual 2D snapshot.  Details concerning
        the phase-dependence of microturbulent velocities of \ion{Fe}{i} and
        \ion{Fe}{ii} lines as a function of the photometric phase in Cepheids
        will be presented in forthcoming papers.
        \subsection{Atomic model and non-LTE line formation}
        \label{methods_nlte}
        
        \citet{2018arXiv180310531A} presented a new oxygen model atom that
        incorporates inelastic O+H collisional rate coefficients {based on the
          asymptotic two-electron model of \citet{2018A&A...610A..57B} and the
          free electron model of \citet{1991JPhB...24L.127K}.  The}
        MPI-parallelised 3D non-LTE radiative transfer code \textsc{Balder}
        \citep{2018A&A...615A.139A}, a custom version of
        \textsc{MULTI3D} \citep{2009ASPC..415...87L}, was used in this study.
        The code solves the statistical equilibrium equation taking into
        account {radiative and collisional bound-bound and bound-free
          transitions} to get level populations {and also to calculate the
          final emergent spectrum.  Line formation calculations were performed
          on $705$~snapshots of M9, spanning $10$~cycles, and $326$~snapshots
          of M3, spanning 14~cycles.}
        
        {The  LTE and  non-LTE calculations on the 
                snapshots of the 2D models
                (\sect{methods_atmos})
                and the  LTE and  non-LTE calculations across
                the grid of 1D models
                (\sect{methods_1d}) proceeded in mainly the same way.
                For the 1D calculations, a depth-independent 
                microturbulent velocity $\xi_\mathrm{t}$~was introduced,
                and was varied between 
                $0.0$ and $4.0\,\kmos$~with a step $1.0\,\kmos$.
                Calculations were performed for several oxygen abundances
                $-0.8<\mathrm{[O/Fe]}<+0.8$~with a step $0.4\,\dex$.}
        \subsection{Abundance corrections}
        \label{method_abcor}
        
        In order to compare the analysis using multi-dimensional model atmospheres with the
        classical 1D approach we can  calculate abundance corrections and
        abundance  errors.  
        {Given a measured LTE abundance,
                we define the abundance correction as} the abundance
        difference compared to LTE,
\begin{IEEEeqnarray}{rCl}
        \Delta^\mathrm{NLTE}_\mathrm{LTE}
        \left(\log\, \epsilon^\mathrm{LTE}\right)&=&
        \log\, \epsilon^\mathrm{NLTE} - \log\, \epsilon^\mathrm{LTE}\,,
\end{IEEEeqnarray}
such that the non-LTE equivalent width matches
                the LTE equivalent width.
        \section{Results}
        \label{results}
        
        \subsection{Departure coefficients}
        \label{results_b}
        
        \begin{figure}
                \includegraphics[width=9cm, scale=1.0] {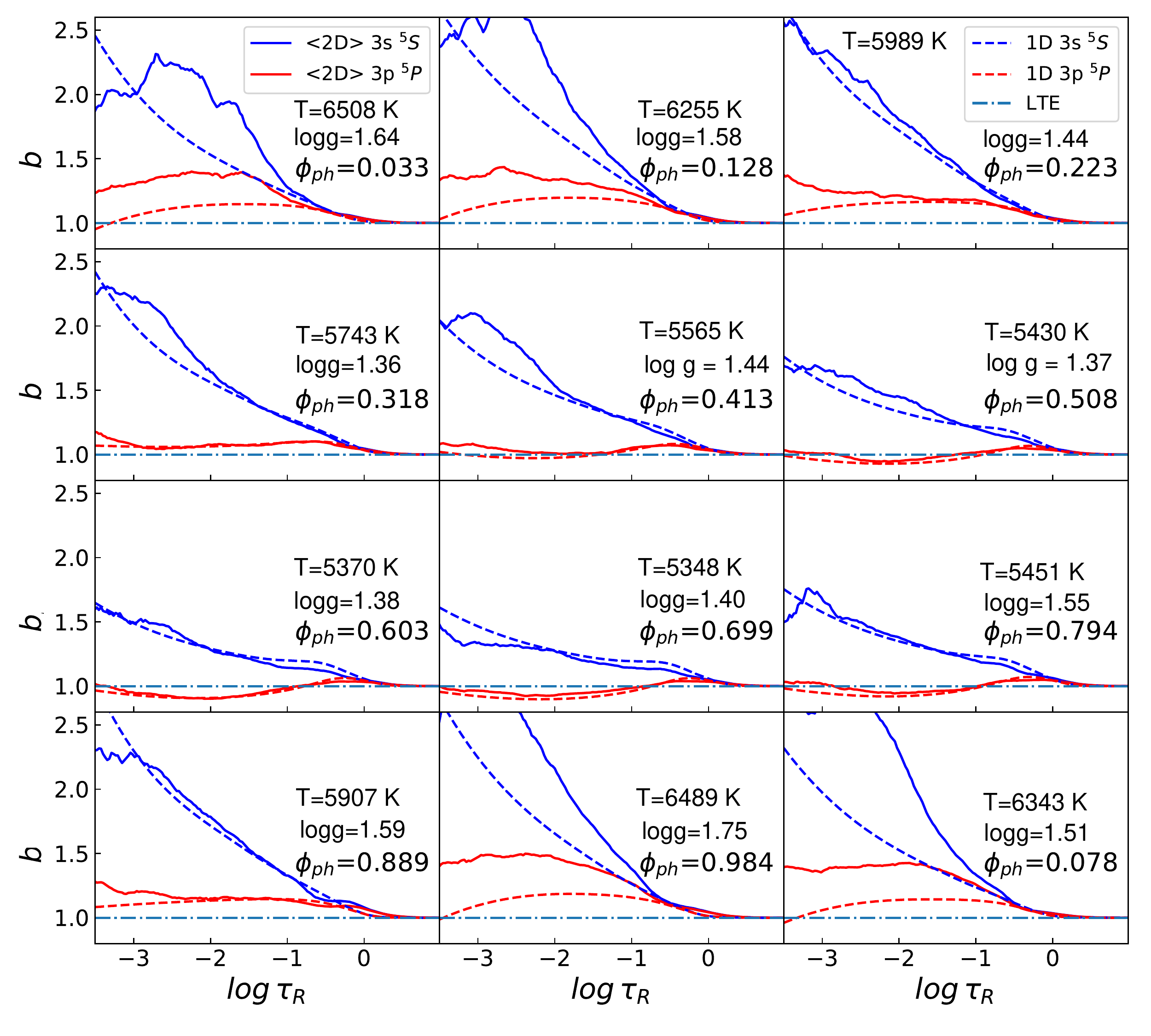}
                \caption{Plots of  1D $b^\mathrm{1D}(\tau_\mathrm{R})$  and
                        median  2D  $b^\mathrm{2D}(\tau_\mathrm{R})$  departures  
                        as a function 
                        of $\tau_\mathrm{R}$ for the lower \lowerlev\  (blue) 
                        and upper \upperlev\ (red) 
                        levels  (dashed  and solid  lines, respectively) 
                        for  a particular  cycle.  
                        The median  2D  departures are 
                        results of the horizontal average along the surfaces
                        $\tau_\mathrm{R}=\mathrm{const}$. 
                        Departures from LTE  $b^\mathrm{1D}(\tau_\mathrm{R})$  are 
                        based  on  interpolation of results of  calculations using  1D hydrostatic models  by fixing 
                        $\Big[ \teff,\lggeff, \xi_\mathrm{t}
                        \Big]^\mathrm{2D}$. }
                \label{mean_departs}%
        \end{figure}

        In order to understand the behaviour  of the  abundance corrections $\Delta^\mathrm{2D\,NLTE}_\mathrm{1D\,NLTE}$ and
        $\Delta^\mathrm{2D\,NLTE}_\mathrm{1D\,LTE}$,  we can look at the variation of
        the departure coefficients,
        $b(\tau_\mathrm{R})=n_\mathrm{NLTE}/n_\mathrm{LTE}$, for the 1D hydrostatic
        models and 2D models. For the first, $b^\mathrm{1D}$  values were calculated
        for the lower \lowerlev\ and upper \upperlev\ levels  of the \ion{O}{i} 777 nm
        triplet in the 4D parameter space $\tau_\mathrm{R}, \teff, \lggeff,
        \xi_\mathrm{t}$.  For each combination $\Big[ \teff,\lggeff,
        \xi_\mathrm{t} \Big]^\mathrm{2D}$  a cubic interpolation that splits the
        multi-dimensional case into a sequence of 1D interpolations was performed on
        the grid of 1D models to obtain departures from LTE as functions of the Rosseland
        optical depth. 
        
         In the 2D models -- due to the horizontal inhomogeneities produced by
         convection -- departures from LTE $b(\tau_\mathrm{R})$ are different for
         in- and outflows   \citep[e.g.][]{2016MNRAS.455.3735A}.  In order to
         compare departures from LTE for 2D models with the 1D case,
         $b^\mathrm{2D}$ values for each individual 2D snapshot were horizontally
         averaged over surfaces of constant optical depth.  Both 2D
         models predict a similar dependence of the abundance corrections on
         effective temperature. Since they are similar we focused on the nine-day
         model, and calculated median 2D departures from LTE.

        We plot the departure coefficients in \fig{mean_departs}.  Namely, the
        median 2D departures from LTE as a function of $\tau_\mathrm{R}$ and
        photometric phase for the lower and upper levels are shown by solid
        blue and red lines, respectively.  One-dimensional interpolated
        $b^\mathrm{1D}$ departures from LTE are shown by dashed lines using
        the same colours for the levels.  {Qualitatively, the departures from LTE of  1D and 2D  models show a similar dependence on the photometric phase.} The lower and upper levels of the \trip~triplet,
        namely  \lowerlev~and \upperlev, are almost always overpopulated
        during the pulsating cycle.  At photometric phases
        $\phi_\mathrm{ph}\approx 0.3 \ldots 0.8$ the lower and upper levels
        are less underpopulated compared to the rest of the phases, due to the
        minimum on the light curve and effective temperature.  Quantitatively,
        for this range of phases 1D and 2D models predict  very similar
        optical depth dependences of the departures from LTE.  As a result, the
        abundance correction $\Delta^\mathrm{2D\,NLTE}_\mathrm{1D\,NLTE}$ is
        small. Moreover, the 2D non-LTE versus 1D LTE abundance correction
        turns out to be smallest, as shown in Sect.~\ref{results_2d1d}.
        \subsection{\trip~triplet profiles}
        \label{results_profiles}

        \begin{figure*}
        \includegraphics[scale=0.37] {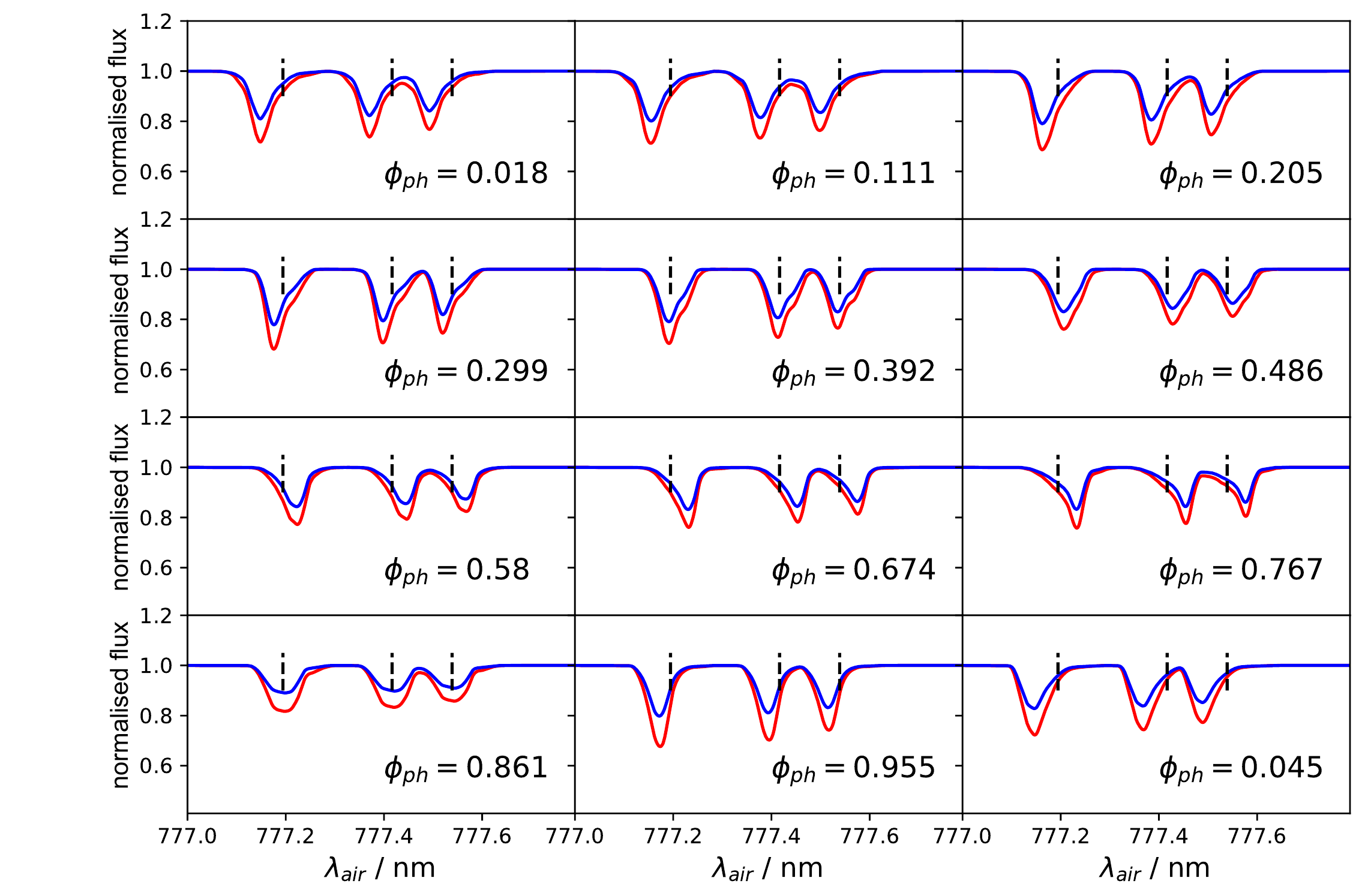}
        \includegraphics[scale=0.37] {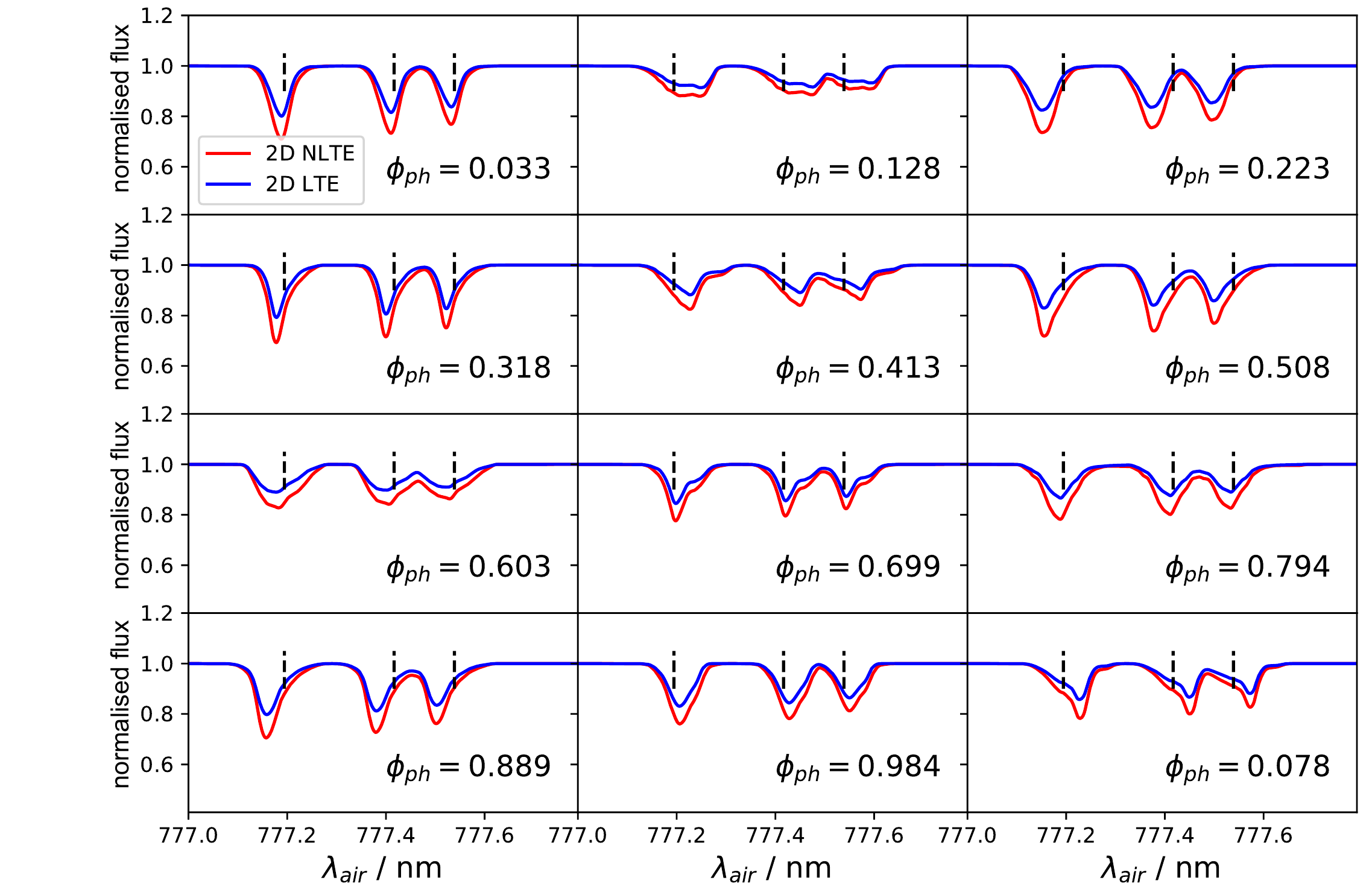}
                \caption{Variation in the \trip~triplet during 
                        a particular pulsational cycle for LTE (blue) and non-LTE (red) cases 
            calculated for the three-day M3  Cepheid model (left panel)
            and for the none-day M9  Cepheid model (right panel).
            The dashed vertical lines indicate the lines' rest wavelengths.}
                \label{line_prof}%
        \end{figure*}
        
        \begin{figure*}
                \includegraphics[width=9cm,scale=0.7] {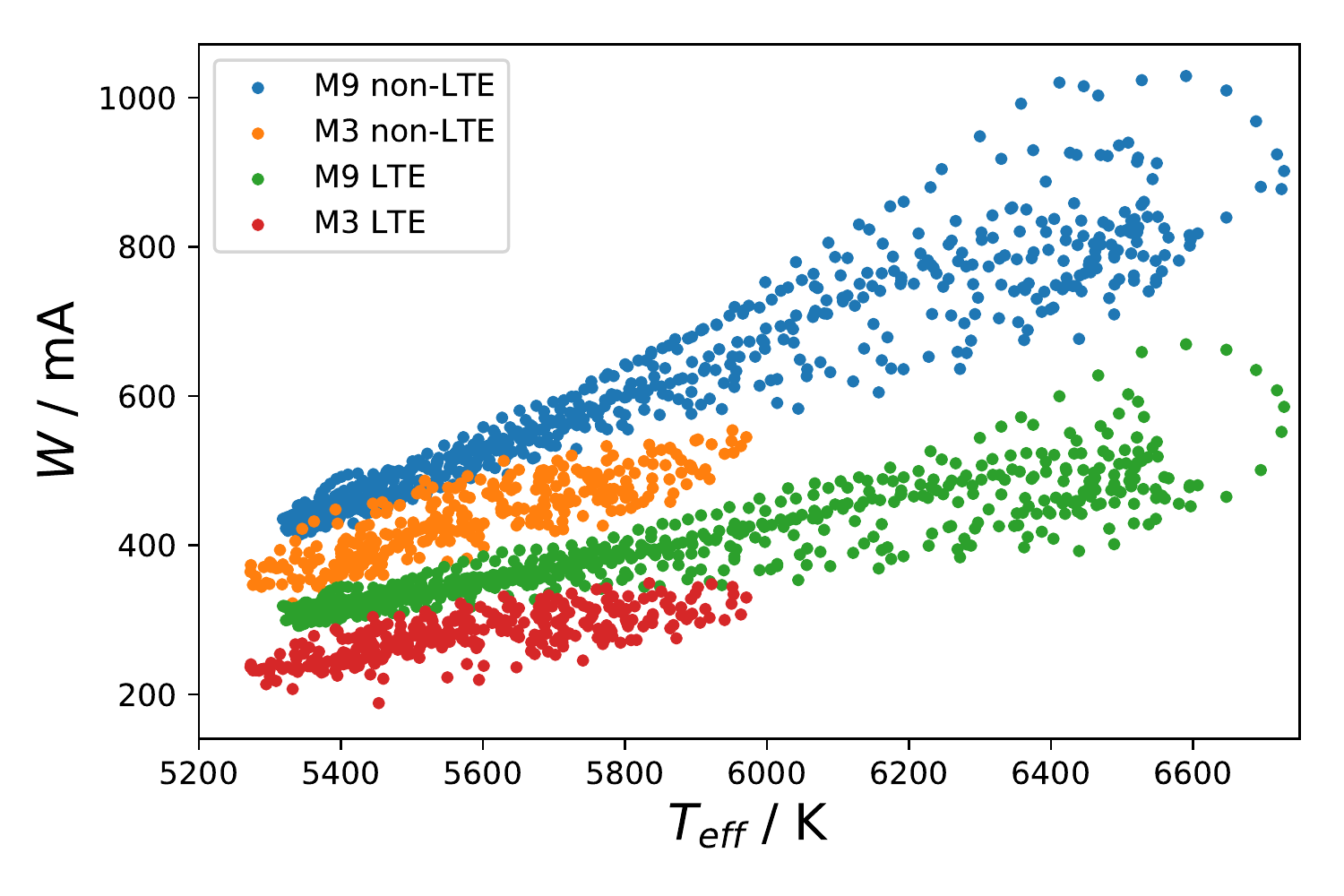}\includegraphics[width=9cm,scale=0.7] {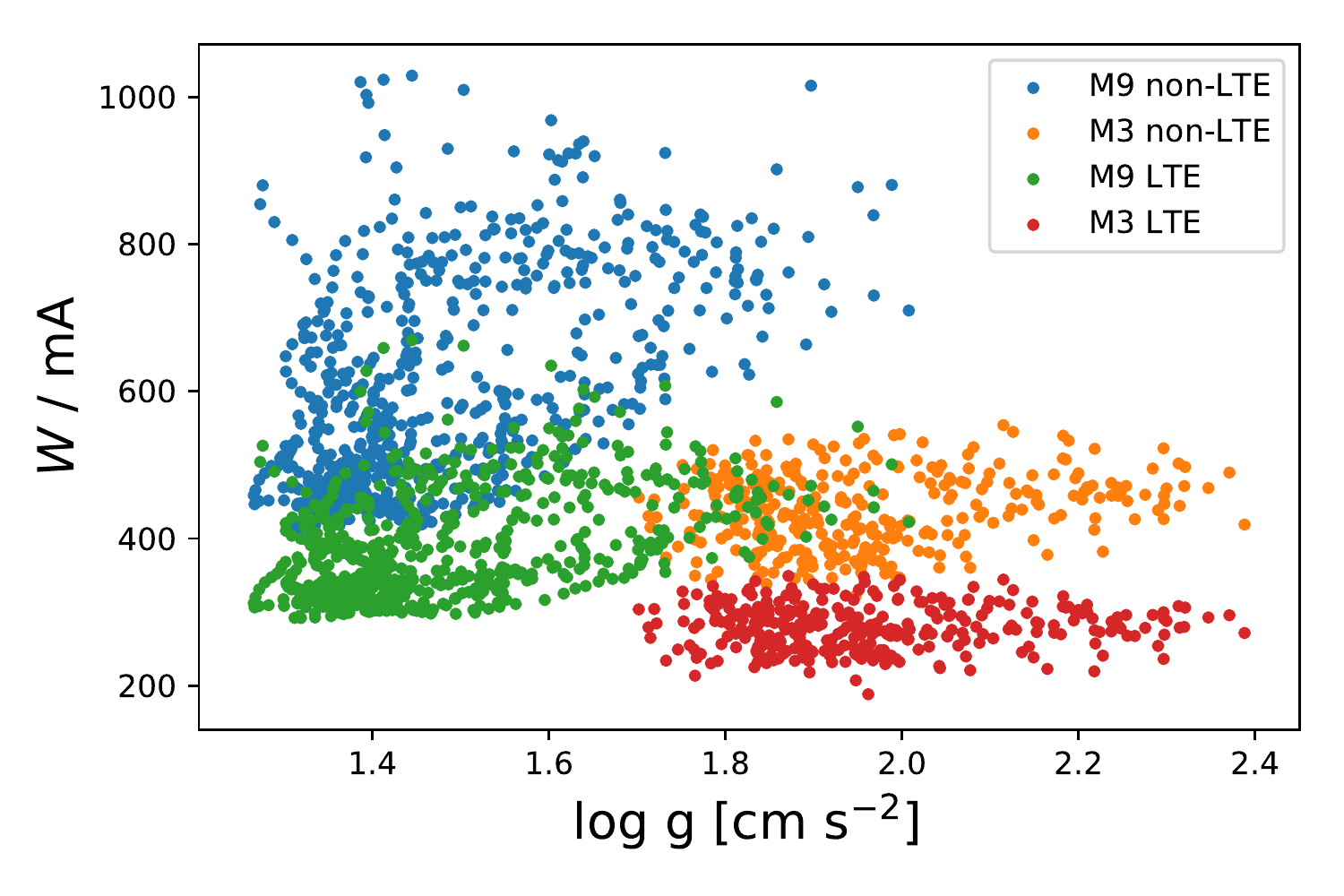}
        \caption{Equivalent widths
                for  the  \trip~triplet integrated over all components  as
                functions of effective temperature and effective surface gravity
        for
                $\mathrm{[O/Fe]}=-0.4,0.0,+0.4$ and models M3 and M9.}
        \label{eqwidths_2d}%
\end{figure*}

        {We illustrate the temporal evolution (over a pulsation cycle) of the
          normalised emergent LTE and non-LTE \trip~triplet flux profiles} in
        \fig{line_prof} for M3 and M9.
        The line profiles are shown without adding instrumental
        broadening or noise. This allows us to identify the occasional
        multi-component structure of the line profiles, for example at photometric
        phases $\phi_\mathrm{ph}=0.299$ and $0.318$ for M3 and M9,
        respectively.  Overall, expansion or compression  produce
        asymmetric line profiles that are typical for pulsating stars
        \citep{1919PNAS....5..417S}.
        


In \fig{eqwidths_2d} we show the equivalent widths as a function
of the effective temperature and the surface gravity, for both
M3 and M9 models. Generally, the non-LTE equivalent widths are larger than the
LTE values, as also seen in \fig{line_prof}.
The main driver of the non-LTE effect is photon losses in the 
\ion{O}{I} $777$\,nm~triplet itself
\citep[e.g.][]{2003A&A...402..343T}, which leads to a
strengthening of the triplet.

{Figure}~\ref{eqwidths_2d} also shows that the equivalent widths
have a strong dependence on the effective temperatures,
and a weaker dependence on the surface gravity.
The models span a wide range of effective temperatures,
and a relatively narrow range of surface gravities
(\fig{1d_grid}): at fixed $\lgg$, $\teff$~spans up to around 
$1000\,\mathrm{K}$, corresponding to around a 
$0.3$~to $0.5\,\dex$~change in line strength;
instead, at fixed $\teff$, 
$\lgg$~spans up to around $0.4\,\dex$, corresponding 
to a less than $0.04\,\dex$~change in line strength, based on
the 1D model atmospheres.
In other words, in this context the line strengths are more
sensitive to the effective temperature than to the surface gravity,
as expected for lines of high excitation potential
from a neutral species.

        \subsection{Abundance differences: 2D non-LTE versus 2D LTE}
        \label{results_nlte}
        
        \begin{figure*}
                \includegraphics[width=9cm, scale=0.7] {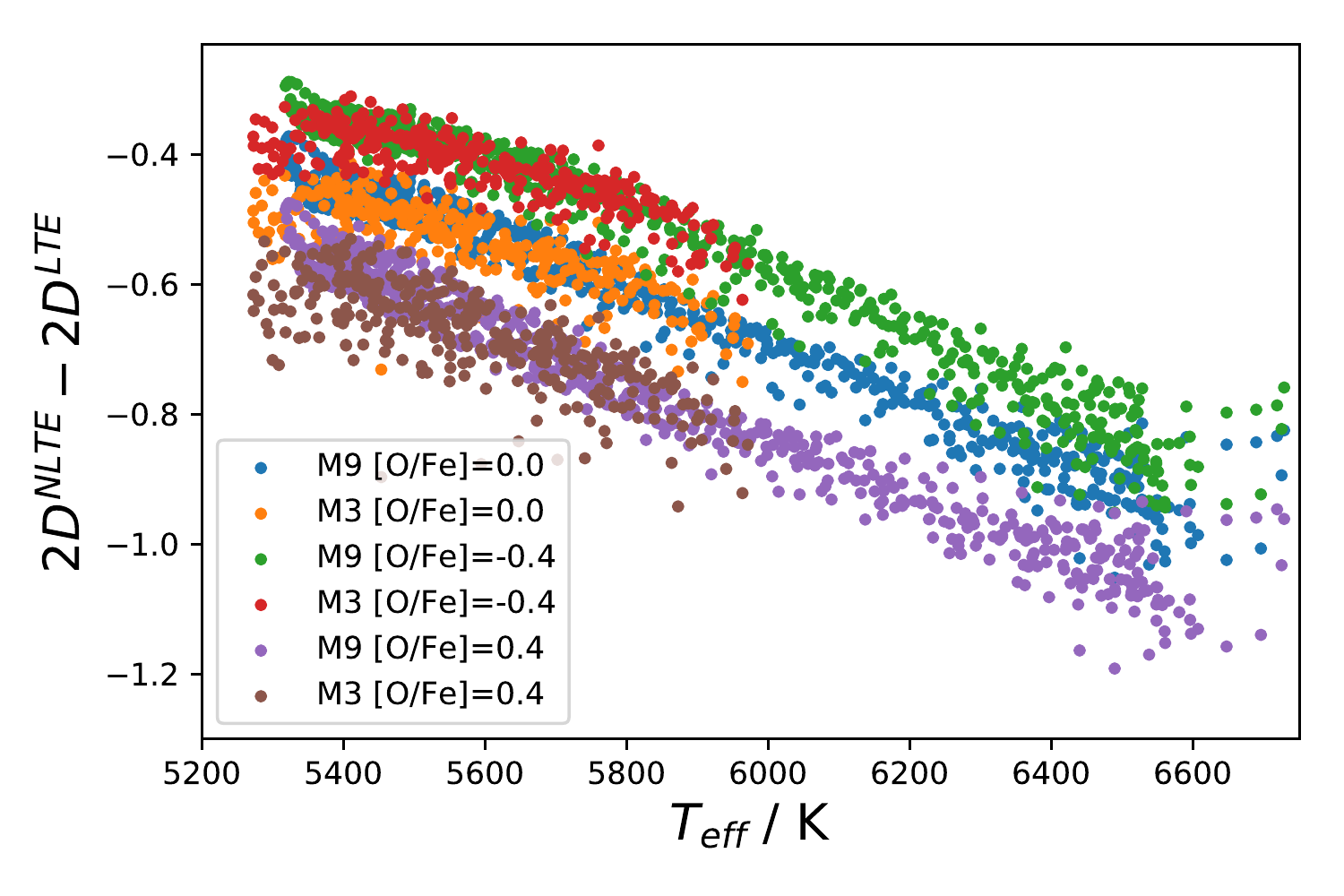}\includegraphics[width=9cm, scale=0.7] {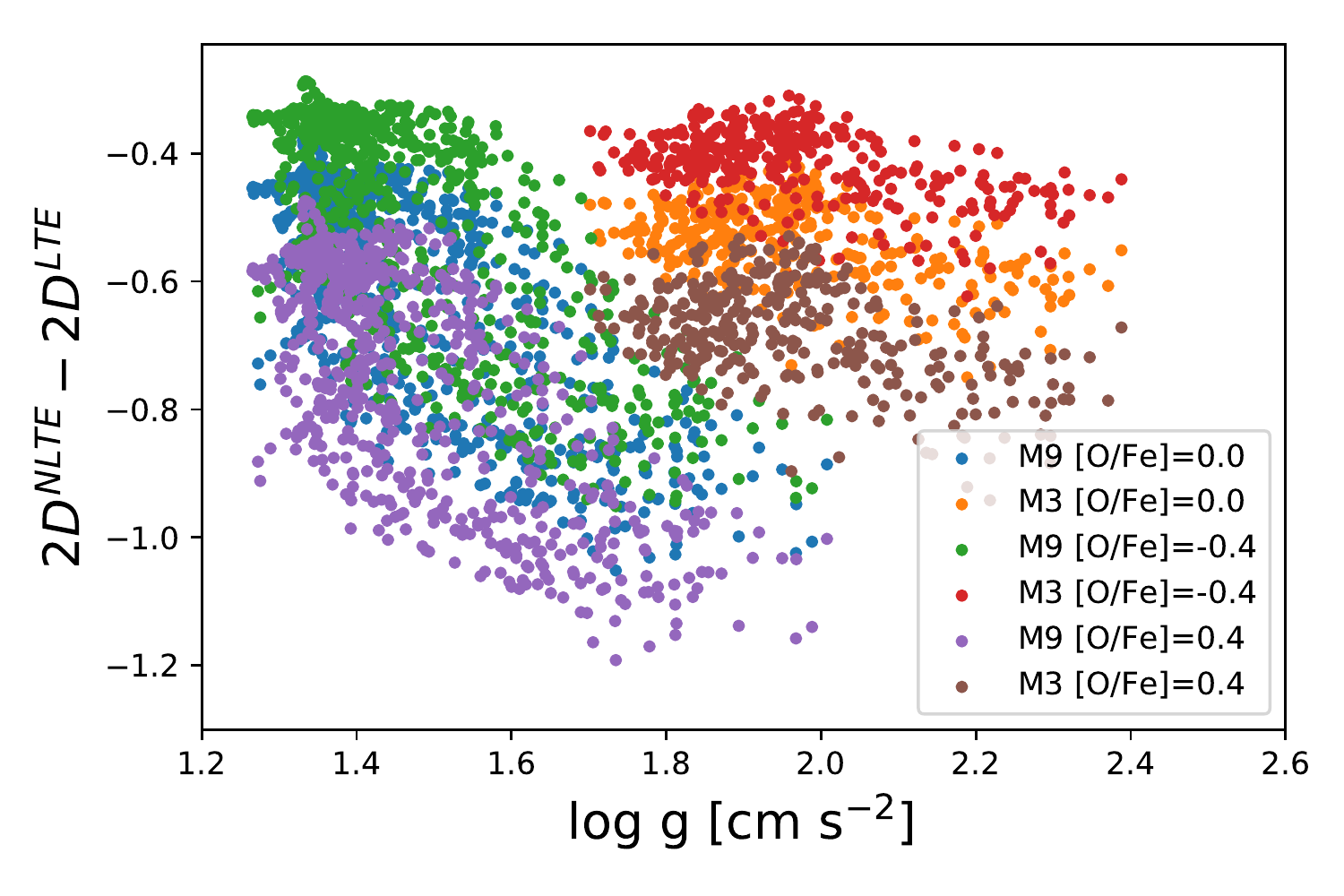}
                \caption{Plots of 2D NLTE vs 2D LTE abundance corrections 
                        for  the  \trip~triplet integrated over all components  as
                        functions of effective temperature and effective surface gravity
            for $\mathrm{[O/Fe]}=-0.4,0.0,+0.4$ and models M3 and M9.}
                \label{ab_err_2d}%
        \end{figure*}
        
In \fig{ab_err_2d} we illustrate
the 2D LTE versus 2D non-LTE abundance corrections
as functions of effective temperature and
surface gravity for the 2D models. There is a larger spread in the 
abundance corrections at fixed surface gravity
than at fixed effective temperature.
Furthermore the M3 and M9 models are clearly offset
in abundance corrections versus surface gravity
space, whereas they overlap in
in abundance corrections versus effective temperature space.
This can be understood from our discussion in \sect{results_profiles}:
the non-LTE effect is one
of photon losses in the lines themselves, and
the triplet is more sensitive to the effective temperature
than to the surface gravity.

{Figure~\ref{ab_err_2d}} shows that the non-LTE effects are
the most severe at higher effective temperatures.
This happens close to the phases of maximum light;
as seen in  \fig{line_prof}, at these phases
the non-LTE line depths are roughly twice as
large as the LTE values. 
The M3 model has smaller line depths ratios between the LTE and non-LTE cases
because at maximum light it has a smaller effective temperature
(and thus weaker \ion{O}{I} $777$\,nm lines and fewer photon losses).
At minimum
light (low effective temperature), the M3 and M9 models have 
similar line depth ratios in both LTE and non-LTE,
with less severe non-LTE effects. Hence, there is a modulation of the non-LTE
effects over the pulsations.
Nevertheless, even at the phase of 
maximum expansion and start of compression (minimum of the
effective temperature) the 2D LTE versus 2D non-LTE abundance errors are of the
order of $+0.6$~dex, and become larger up to $+1.3$~dex at
$T_\mathrm{eff}\approx6700$ K. In order to obtain unbiased estimates of the oxygen
abundance in Cepheids (based on the 
\ion{O}{I} $777$\,nm triplet), we have to consider the effects of
departures from LTE.

The scatter of the abundance correction at fixed effective temperature is
related to the impact of the convective noise. It is expected to become weaker
when an average over a larger number of convective elements is performed, for example
in 3D models or by averaging across many cycles. The M3 model shows a larger
scatter due to the stronger impact of the convective component at higher
surface gravity.

        \subsection{Abundance differences: 1D non-LTE versus 1D LTE}
        \label{results_nlte_1d}
\begin{figure}
        \includegraphics[width=9cm, scale=0.7] {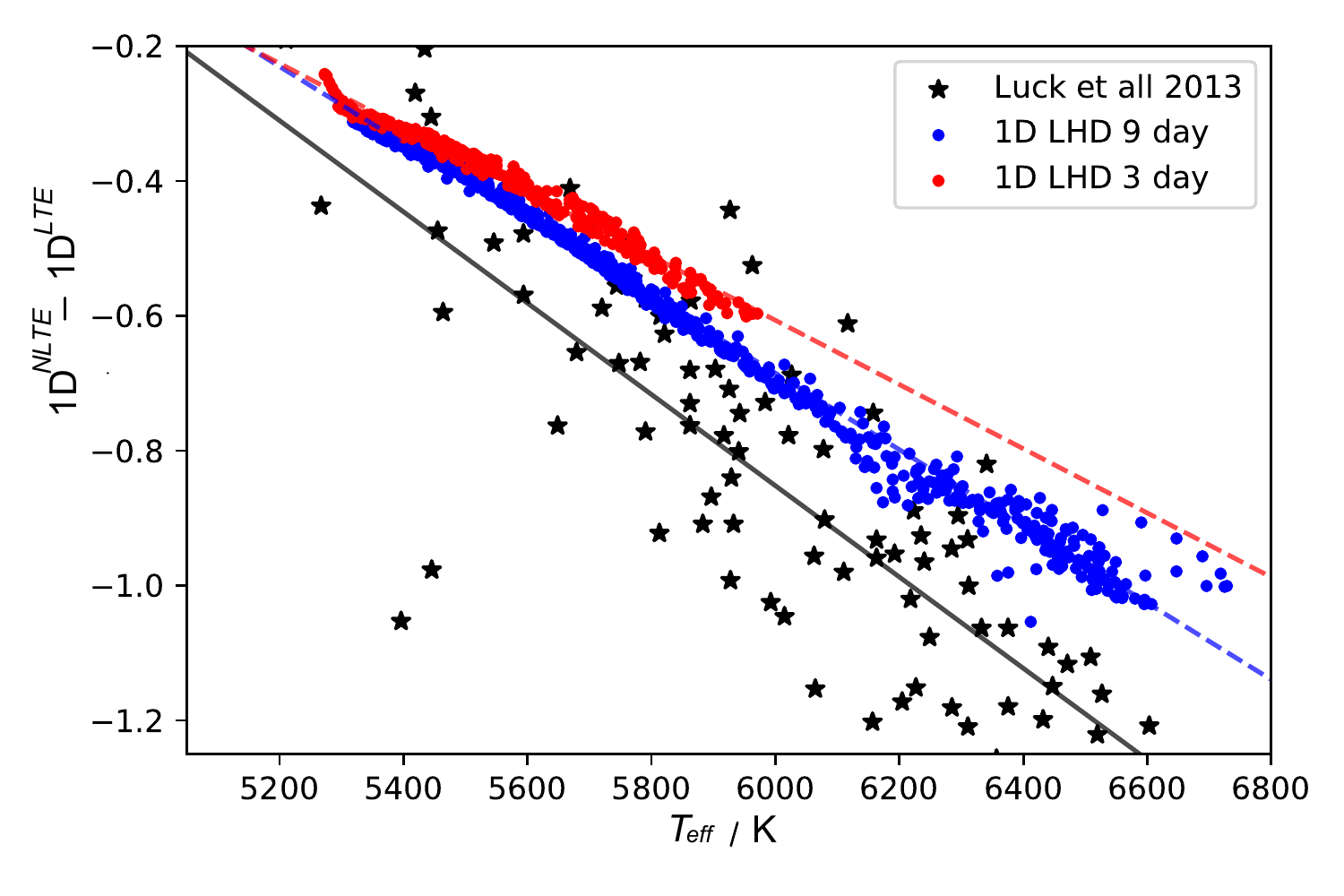}
                \caption{Plot of 1D NLTE vs 1D LTE abundance corrections for the
                  777.4~nm component of the \trip~triplet as a function of
                  effective temperature.  Literature results of
                  \cite{1538-3881-146-1-18} based on 1D ATLAS9 model
                  atmospheres are shown as black stars. Our 1D results are
                  shown as dots taking effective temperatures and gravities
                  from the three-day (red) and nine-day (blue) Cepheid model.}
                \label{ab_corr_1d}%
        \end{figure}

        We illustrate variations in the 1D non-LTE versus 1D LTE abundance
        corrections for the $777.4\,\nm$~component of the \trip~triplet as a
        function of effective temperature for 1D models in \fig{ab_corr_1d}.
        \citet{1538-3881-146-1-18} derived abundance corrections for an
        ensemble of Galactic Cepheids for the \ion{O}{i} $777.4\,\nm$~line as
        function of effective temperature which are shown as black stars in
        Fig.~\ref{ab_corr_1d}.  As  can be seen, their results are
        qualitatively similar to ours; however, the temperature dependence of
        their abundance corrections is steeper.  
        
        This could be due to several
        factors.  First of all, M3 and M9 are not complete stellar
        models. Their pulsational amplitudes and periods depend to some extent
        on the applied geometry and boundary conditions. This leads to a
        different $\teff-\log\,g- \mathrm{Amplitude} - \mathrm{Period}$
        relation with respect to what is expected for real
        Cepheids. Nevertheless, we note here that when focusing on
        spectroscopic properties certain observed spectroscopic features can
        be reproduced by the 2D models \citep{2017A&A...606A.140V}.
        
       Secondly, the surface gravity of our 1D models is based on the
        effective acceleration $\vec{g}_\mathrm{eff}$ of an Lagrangian mass
        element located near optical depth unity in the 2D models. This
        differs from a $\vec{g}_\mathrm{eff}$ adopting the condition of
        ionisation balance, which is traditionally used in classical 1D
        analyses to measure the surface gravity \citep{2018A&A...611A..19V}.
        
        Thirdly, \cite{1538-3881-146-1-18} used 1D hydrostatic ATLAS9 model
        atmospheres \citep{1992IAUS..149..225K}, whereas we used the 1D
        \textsc{LHD} model atmospheres described in \sect{methods_1d}; the use
        of grey radiative transfer has a significant impact on the temperature
        stratification of the \textsc{LHD}~models, and it is possible that
        this may be the reason for most of the differences.  Differences in
        the atomic models may also contribute.

        Additionally, the result of \cite{1538-3881-146-1-18} shows a larger
        vertical scatter for fixed effective temperature.  They explain it by
        the sensitivity of the oxygen abundance to the microturbulent velocity
        and strengths of the line.

        \subsection{Abundance differences: 2D versus 1D}
        \label{results_2d1d}
        
        \begin{figure*}
                \includegraphics[width=19cm, scale=1.0] {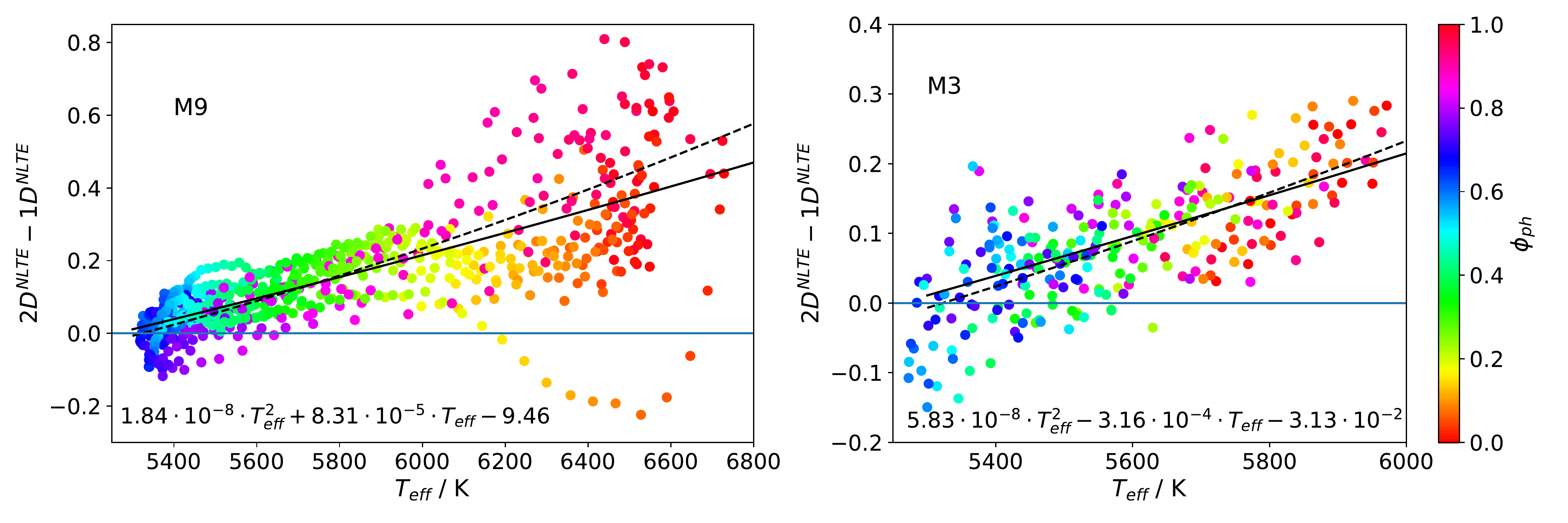}
                \includegraphics[width=19cm, scale=1.0] {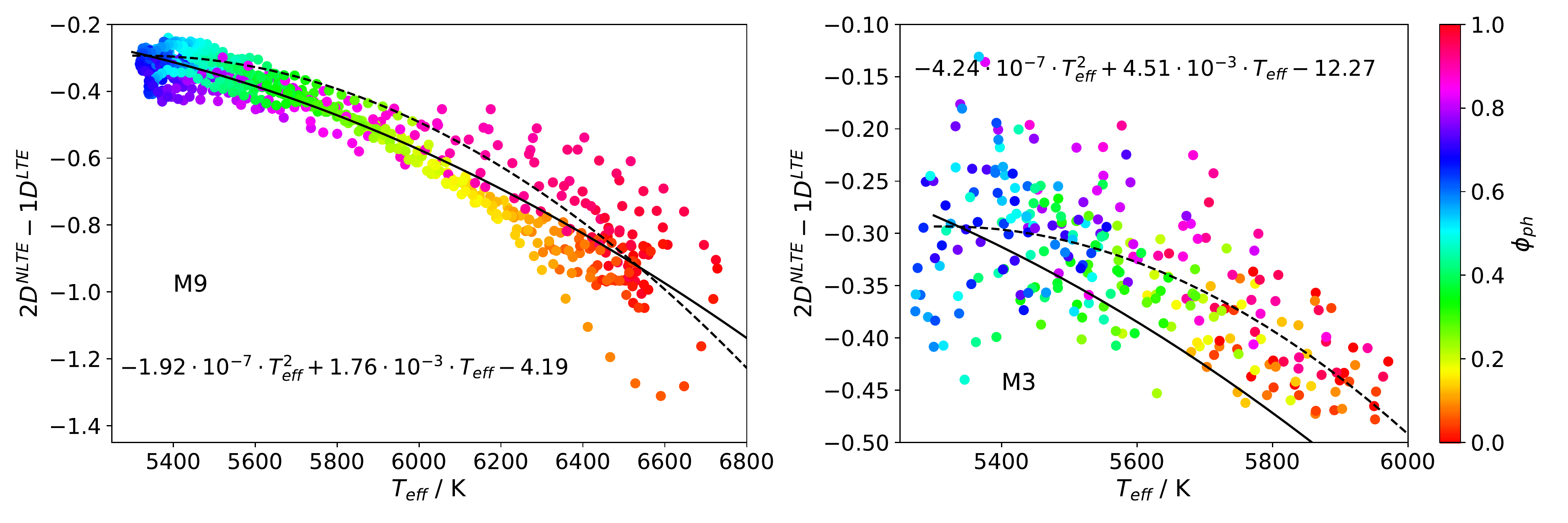}
                \caption{Plot of 2D non-LTE vs 1D non-LTE
            (upper panel) and
            2D non-LTE vs 1D LTE
            (lower panel) abundance corrections 
                        for the \trip~triplet integrated over all  components as 
                        functions of effective temperature for the  
                        nine-day (left panel) and three-day (right panel)
                        Cepheid model.
                        The photometric phase $\phi_\mathrm{ph}$ is indicated by colour. 
                        Solid  and dashed black lines depict the temperature
                        dependence for M9 and M3, respectively, when fitting
                        the numerical data by a parabola.} 
                \label{abundance_corrections_2DNLTE-1D}%
        \end{figure*}
        
        In this section we compare the 2D and 1D approaches to {determine} the optimal
        phases for measuring the \trip~triplet such that the 2D versus 1D
        differences are at a minimum.  To associate a 1D model with a given 2D
        snapshot, we adopt an effective temperature and effective surface
        gravity as described in \sect{methods_atmos}, and a
        microturbulent velocity as described in \sect{methods_vmic}. As stated
        before, this does not exactly correspond to the standard analysis
        procedure where the effective temperature is typically determined from
        line depth ratios \citep{1994PASP..106.1248G,2000A&A...358..587K}, and
        the gravity from the ionisation balance of \ion{Fe}{i} to
        \ion{Fe}{ii}; the microturbulent velocity is obtained from the
        condition of a consistent (often iron) abundance derived for weak and
        strong lines (in LTE).
        
        In our case, for each 2D snapshot the equivalent widths from the grid
        of 1D models were interpolated for the 2D parameters $\left[\teff,
          \lggeff, \xi_\mathrm{t}\right]^\mathrm{2D}$, and the abundance
        corrections and abundance errors were then calculated as stated
        earlier.  The  multi-dimensional interpolation was split into a sequence
        of 1D cubic interpolations. For the 2D non-LTE versus 1D non-LTE
        abundance corrections we illustrate the results of the interpolation
        for M3 and M9 in the upper panel of 
        \fig{abundance_corrections_2DNLTE-1D}. The total
        equivalent widths of the \trip~triplet were taken in the calculations
        here. Analogously, 2D non-LTE versus 1D LTE abundance corrections are
        presented in the lower panel of
        \fig{abundance_corrections_2DNLTE-1D}.
        
        In \fig{abundance_corrections_2DNLTE-1D} models M3 and M9 show
        similar slopes for the 2D non-LTE versus 1D non-LTE abundance
        corrections as functions of effective temperature.  The slope does not
        depend on the surface gravity and is primarily
        controlled by the change of the thermal structure over the pulsational
        cycle. Interestingly, we find that at certain phases a 1D non-LTE
          analysis would provide the same abundance as a 2D non-LTE analysis.
                Qualitatively speaking, these phases are near maximum
                expansion and start of compression.   From these phases,  at higher effective temperatures,  the bias
                grows such that the 1D non-LTE analysis gives abundances that
                are as much as $0.5\,\dex$~too low.
        
        In \fig{abundance_corrections_2DNLTE-1D} models M3 and M9 show
        similar slopes for the 2D non-LTE versus 1D LTE abundance corrections
        as functions of effective temperature.  As with
        the 2D non-LTE versus 1D non-LTE abundance corrections,
        the slope does not depend
        on the surface gravity and is primarily controlled by the change in
        the thermal structure over the pulsational cycle.  Unlike the
        situation for the 2D NLTE versus 1D NLTE corrections, however, we do not
        find any phases at which a 1D LTE analysis would provide the same
        abundance as a 2D non-LTE analysis.  In the best case, at certain
        photometric phases, corresponding to effective temperatures
        $\teff\lesssim5700\,\mathrm{K}$, the abundance corrections are almost
        insensitive to the change in $\teff$ and are around $-0.3\,\dex$,
        ranging from $-0.2\, \dex$ to $-0.4\,\dex$.  For higher effective
        temperatures the abundance corrections become more severe, decreasing
        to $-1.5\,\dex$.
        
        In \fig{abundance_corrections_2DNLTE-1D}, for certain phases 
         the abundance corrections show loop-like structures. As we 
         discussed in Sect.~\ref{methods_atmos}, { for one cycle at maximum 
         light,  $\vec{g}_\mathrm{eff}$ drops by a factor of three while  ${T}_\mathrm{eff}$ }  increases by 
         $130$~K. For this short time interval, the increase in the  
         \trip~triplet  component equivalent widths in LTE and non-LTE 
         cases are primarily due to the sensitivity to the 
         changes in $\lggeff$.  In our calculations,  the relative changes of   LTE and non-LTE  equivalent widths  are
         $\left( \frac{\Delta W}{W} \right) _\mathrm{LTE}\approx 0.4$ and
         $\left( \frac{\Delta W}{W} \right) _\mathrm{NLTE}\approx 0.2$, respectively.  As a
         result, the 2D non-LTE versus 1D LTE abundance error increases  from $-1.35\,\dex$ to
         $-1.1\,\dex$.
        \section{Discussion}
        \label{discussion}
        
        \subsection{Optimal phases for 1D analyses}
        \label{discussion_phase}
        
        \begin{figure*}
        \includegraphics[scale=0.56] {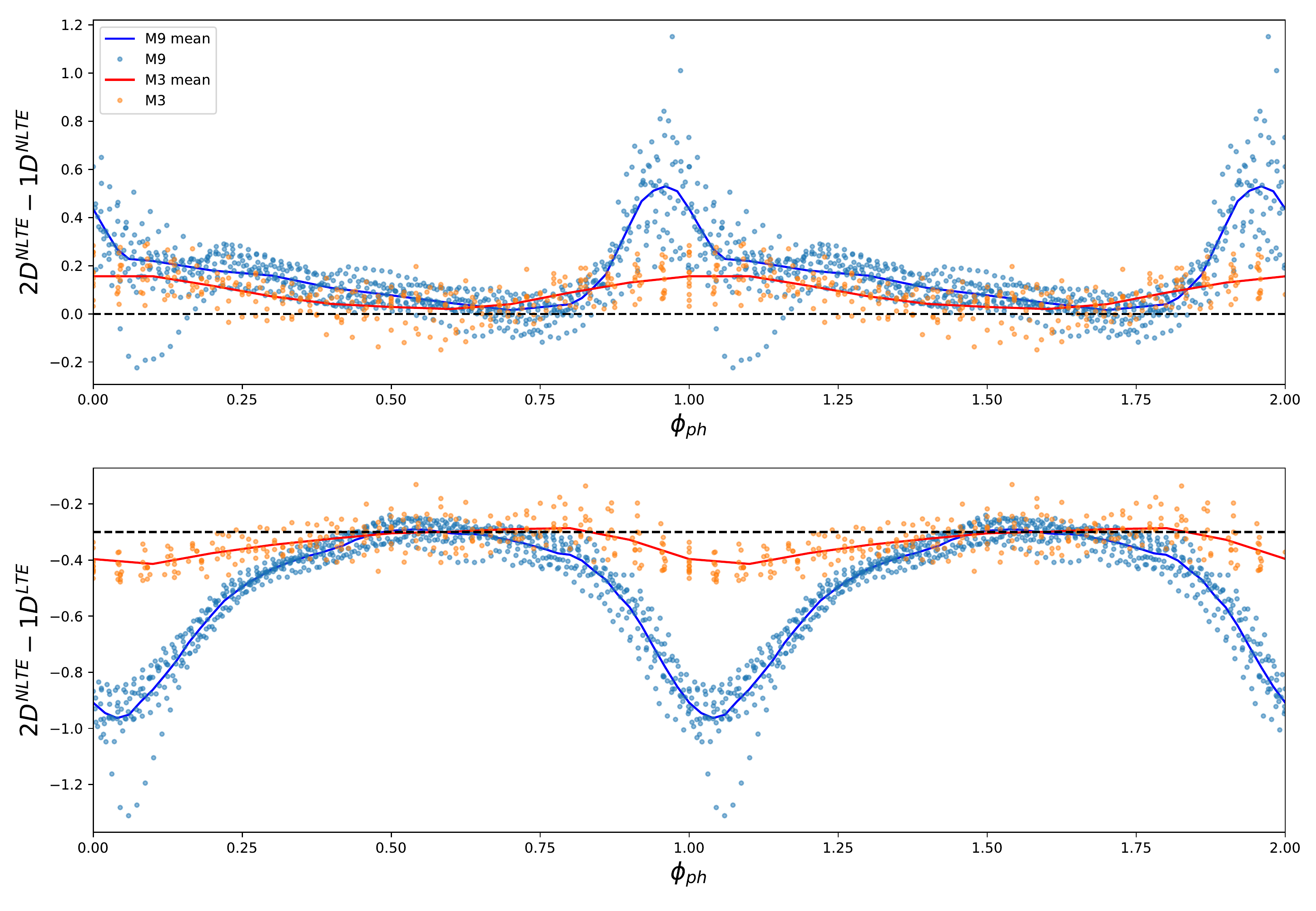}
        \caption{Dependence of the 2D non-LTE vs 1D non-LTE (upper
                  panel) and 2D non-LTE vs 1D LTE (lower panel) abundance
                  corrections for the \trip~triplet integrated over all
                  components as a function of photometric phase for the model
                  M3 (red) and M9 (blue).}
        \label{optimal_phase_2DNLTE-1DNLTE}%
        \end{figure*}

        In practice, spectroscopic analyses of the \trip~triplet are  based on 1D hydrostatic model atmospheres, and
will
        continue to be for the
        foreseeable future.  However, based on the 2D non-LTE abundance
        corrections compared to 1D non-LTE and 1D LTE, we can provide
        recommendations on which phases to target, to obtain the most reliable
        oxygen abundances.  Our recommendation for 1D LTE and 1D non-LTE
        analysis is to avoid phases around maximum light as shown in \fig{optimal_phase_2DNLTE-1DNLTE}.  In a 1D LTE
        analysis,  in the best case we can obtain abundance corrections of the
        order of $-0.3 \pm0.1 \,\dex,$ and in the case of 1D non-LTE analysis $0
        \pm0.1 \,\dex$ at photometric phases $\phi_\mathrm{ph} \approx 0.3
        \ldots 0.8$ independent of period.
        Within these phases, disturbances of the thermal structure and
        convection are not as strong as during maximum compression.  As a
        result, the atmosphere is roughly in hydrostatic equilibrium, and the
        1D models reasonably reproduce the mean thermal structure of the line
        formation region of the 2D model.  If possible, a 1D non-LTE analysis
        based on these phases would be preferable.
        
        \subsection{Implications on Galactic oxygen abundances}
        \label{discussion_gradients}
        
\begin{figure}
\includegraphics[scale=0.6] {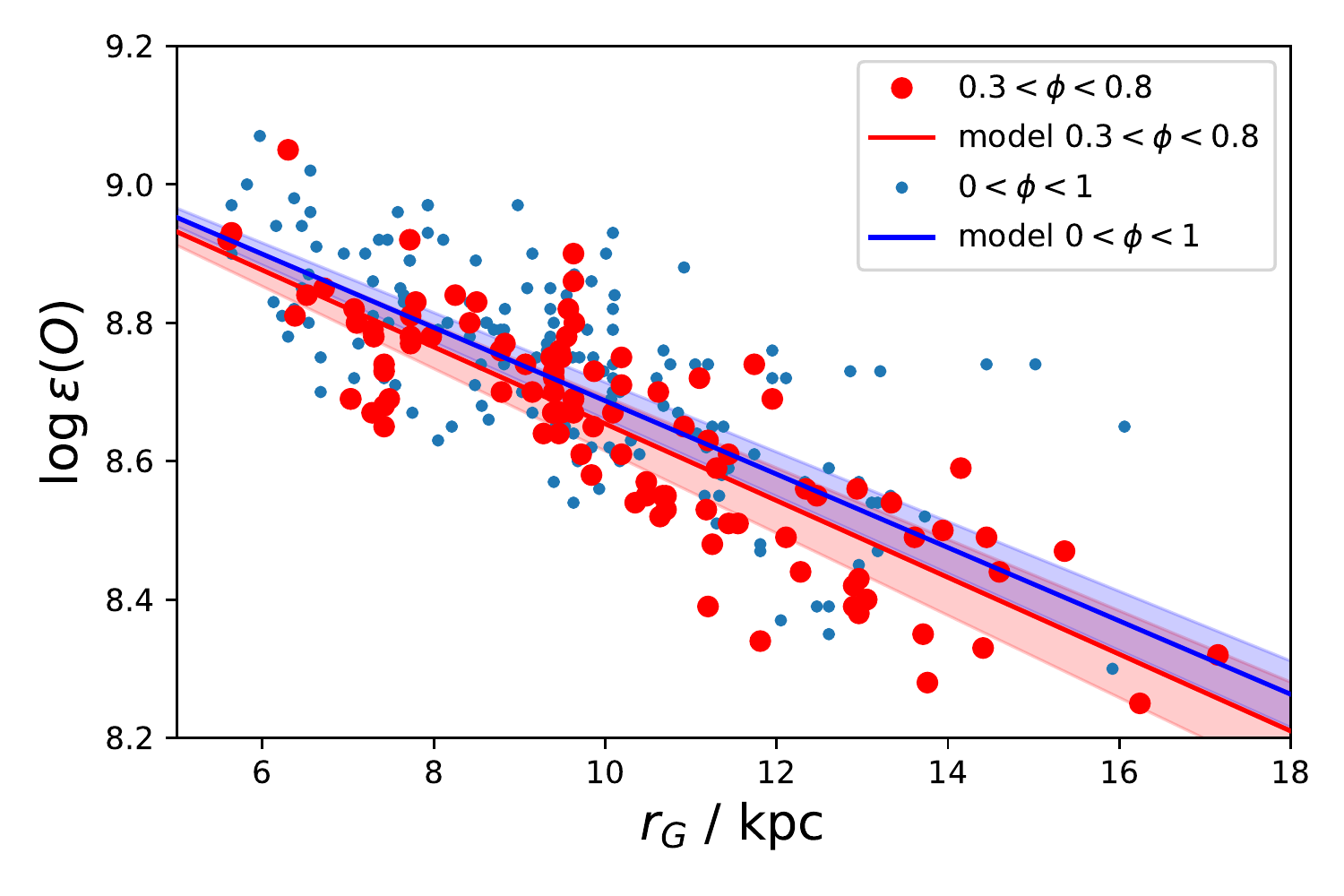}
\caption{Plot of 1D non-LTE oxygen abundances in the Galactic disc, as presented by
  \cite{2014MNRAS.444.3301K}.  Abundances that are based on the analysis of
  spectra taken at all phases and at optimal phases $0.3 <\phi_\mathrm{ph} < 0.8$ are
  shown as light blue dots and red circles, respectively.  Linear fits for
  these two cases including $1\sigma$ uncertainties are presented in red and
  blue {lines}. }
                \label{O_grad}%
         \end{figure}

         
In order to investigate the impact of the phase variation of the 2D non-LTE
versus 1D non-LTE abundance corrections on the Galactic oxygen abundance
gradient, 1D non-LTE oxygen abundances for Galactic Cepheids based on the
\ion{O}{I} 777\,nm triplet were taken from \cite{2014MNRAS.444.3301K}.  Using
a linear fit we measured the gradient for two cases: (i) spectra were taken at
all  phases and (ii) at optimal phases $0.3 <\phi_\mathrm{ph} < 0.8$ for which a 1D
non-LTE analysis should provide largely unbiased abundances according to our
predictions.  The gradients are presented in Fig.~\ref{O_grad} as blue and red
lines, respectively, including $1\sigma$ uncertainties. For the data obtained
at all phases the slope is $-0.053 \pm 0.003$ dex$\cdot$kpc$^{-1}$, whereas
with data at optimal phases we get a slightly steeper gradient $-0.056 \pm
0.004$ dex$\cdot$kpc$^{-1}$. However, the change in  slope is within the
uncertainties.
  
It is also possible to apply the 2D non-LTE versus 1D non-LTE abundance
corrections directly to a 1D analysis; however, this requires 2D models with
different oxygen abundances and metallicites, as well as pulsation properties,
in order to cover the observed range of parameters.  Recent multi-phase
analyses of Galactic Cepheids by \cite{2018AJ....156..171L} and \cite{
  2018A&A...616A..82P} using 1D model atmospheres show some dependence of
elemental abundances on the photometric phase.  In particular, a dependence is
seen in the iron abundances, and they are possibly an indication of the
failure of 1D hydrostatic models to represent the 3D hydrodynamic nature of
a pulsating Cepheid atmosphere.  For oxygen, the phase dependence
should be stronger for the the \ion{O}{I} 777\,nm triplet owing to its
temperature sensitivity, which is amplified by temperature-sensitive 3D
non-LTE effects.  We expect  the dependences to be  stronger in Cepheids with higher
mean effective temperatures, to be  more metal-rich, and to have longer periods.
   
        \section{Conclusion}
        \label{conclusion}

        In this paper for the first time we checked the validity of the
        standard non-LTE spectroscopic analysis of oxygen in the context of
        Cepheid variables. To achieve this goal, we carried out 2D RHD
        simulations pulsating Cepheid atmospheres with three- and nine-day periods. In
        post-processing, we performed 2D non-LTE radiative transfer
        calculations for atomic oxygen.  We compared the 2D results to those
        based on 1D hydrostatic model atmospheres.  We find a strong
        modulation of the non-LTE effects in the oxygen 777 nm triplet lines
        during the pulsations phase, mostly due to the change of the effective
        temperature.  The dependence of the abundance differences
        $\Delta^\mathrm{2D\,NLTE}_\mathrm{2D\,LTE}$ based on the 2D Cepheid
        models is different with respect to the analysis of
        \cite{1538-3881-146-1-18} obtained with the non-LTE 1D hydrostatic
        approach.  We confirmed that 1D LTE analyses of the \ion{O}{I} 777\,nm
        triplet cannot provide unbiased estimates of the oxygen abundance at
        any phase. This also implies that 2D effects do not cancel out the non-LTE
        effects.
        
        The 1D non-LTE analysis of the \ion{O}{i} 777 nm triplet lines can
        provide an unbiased estimate of the oxygen abundance only for a limited
        range of photometric phases. For the analysis using the whole oxygen
        777 nm triplet, it is desirable to analyse spectra taken at photometric
        phases {$\phi_\mathrm{ph}\approx0.3\ldots0.8,$} independent of
        period, when the pulsating atmosphere is closer to hydrostatic
        conditions. {For these phases} 1D as well as 2D non-LTE calculations predict a
        similar dependence of the departure coefficients for the lower and
        upper level on the optical depth.
        
In this work we presented a first step towards more reliable spectroscopic
analyses of Cepheid variables.  Nevertheless, there are a number of ways
in which the models still have to be improved.  The model atmospheres
themselves are 2D: for non-variable stars there is evidence that 
full 3D simulations are required
to generate the correct velocity fields and temperature and pressure
stratifications \citep{2000A&A...359..669A}; we expect that similar
problems may exist here.  In particular, with full 3D simulations we would expect  less influence of the stochastic convective noise on pulsating properties and  thermal structure of the atmosphere.  It reduces   cycle-to-cycle  variations of the light curve  {and ultimately scatter,  which we see near the maximum light phases}   in   LTE  versus  non-LTE abundance corrections at fixed effective temperature.  Additional avenues of work involve constructing
the atmospheres using non-grey radiative transfer, perhaps by
employing opacity bins. 
Another important step is  to broaden the model basis to better
sample the $T_\mathrm{eff}$ -- $\log\,g$ -- [Fe/H] -- Amplitude -- 
Period space.  This is a necessary step, before 2D non-LTE (or 
3D non-LTE) can be directly applied to the literature 1D LTE abundances.
        \begin{acknowledgements}
                VV, HGL, and BL acknowledge financial support from the Sonderforschungsbereich SFB\,881
                `The Milky Way System' (subprojects A4, A5) of the German Research
                Foundation (DFG) and MPIA IT department for computer resources. 
                AMA acknowledges funds from the Alexander von Humboldt Foundation in the
                framework of the Sofja Kovalevskaja Award endowed by the Federal Ministry of
                Education and Research. This  work
                was  also  made  possible  by  the  open-source  codes
                \textsc{matplotlib} \citep{2007CSE.....9...90H}, \textsc{numpy}
                \citep{5725236}, \textsc{scipy} \citep{scipy}, and \textsc{pandas} \citep{
                        mckinney-proc-scipy-2010}. 
        \end{acknowledgements}
        \bibliographystyle{aa} 
        \bibliography{bibliography.bib}

\end{document}